\documentclass[twocolumn,showpacs,preprintnumbers,amsmath,amssymb]{revtex4}
\usepackage{graphicx}
\usepackage{dcolumn}
\usepackage{bm}
\begin{document}
\title{Ground-state energy and frustration of the Sherrington-Kirkpatrick model and related models}

\author{S. Kobe}
\email{kobe@theory.phy.tu-dresden.de}
\affiliation{Technische Universit\"{a}t Dresden,
Institut f\"{u}r Theoretische Physik, D-01062 Dresden, Germany}

\date{\today}
\begin{abstract}
Exact ground states are calculated for the Sherrington-Kirkpatrick (SK) spin-glass containing up to $N=90$ spins. A ground-state energy per spin $e^{\infty}_0 =  - 0.7637 \pm 0.0004$ is found from the $N$ dependence of the misfit parameter, which is a measure of frustration of the
system. The results are compared with those of two related models, which can be introduced by replacing all interactions of the SK model by
ferromagnetic or antiferromagnetic ones of the same strengths. A parameter $x$ is introduced, which describes the fraction of antiferromagnetic interactions in these types of models. From the $x$ dependence of finite models it is concluded, that the the SK model $(x=0.5)$
assigns a transition between a ferromagnetic state $(x<0.5)$ to a spin-glass state $(x>0.5)$.
\end{abstract}

\maketitle

The Sherrington-Kirkpatrick (SK) model \cite{SK} was introduced to describe the spin-glass phase, which is relevant to understand the physics of glassy matter. Its importance arises from the possibility to get analytical results (see e.g. \cite{binyoung}), but it is also an object of mathematical interest \cite{tala}. Currently, the question is controversially debated, whether spin glasses are 'still complex' after some decades of intensive studies (\cite{steinrev}, \cite{franz}, \cite{plefka} and references therein). Another problem concerns if the results of the SK model are valid also for more realistic short-range spin glasses \cite{hed}. In this context, the question of the finite-size scaling is widely discussed.

The problem of finding the exact ground state of most of the Ising spin-glass models belongs to the class of NP-hard problems, i.e. no algorithm is known which finds the optimum in polynomial time \cite{hartrie}. Therefore, in the past, investigations of the size dependence of the ground states often were based on approximations. Very recently, attention has been paid to the problem of fluctuations of spin-glass ground-state energies \cite{and}, \cite{pal}.

In this letter a numerical investigation of exact ground states up to $N=90$ spins is presented, which are obtained with the branch-and-bound algorithm
of discrete nonlinear optimization \cite{koha}, \cite{hadako}, see also \cite{korev}.

Starting from the Hamiltonian
 \begin{eqnarray}
 \mathcal H = - \frac{1}{\sqrt{N}} \sum \limits _{1 \le i < j \le N} J_{ij}S_iS_j  \;,
 \end{eqnarray}
the SK model is described
using $J_{ij} = G_{ij}$, where the independent identically distributed random numbers $G_{ij}$
are chosen from a Gaussian distribution with zero means and variance one. The chosen scaling in (1) secures that the
spin-glass ground-state energy is an extensive quantity. With $J_{ij} = |G_{ij}|$
a fully connected ferromagnetic system with the ground-state energy per spin $e^{id}_{0} = (N-1)/(2\pi N)^{1/2}$
is obtained. Because $e^{id}_{0}$ scales like $N^{1/2}$, the ferromagnetic model should be considered in the following as a
limiting case of the SK model irrespective of its physical meaning.

Another limiting case is the fully connected antiferromagnetic (afm) system:
$J_{ij} = - |G_{ij}|$. A further interpolation of the described models can be obtained considering
a distribution of $J_{ij}$ according
\begin{eqnarray}
P(J_{ij}) =x \delta(J_{ij} + |G_{ij}|)+(1-x) \delta (J_{ij} - |G_{ij}|).
\end{eqnarray}
Starting from the SK model ($x = \frac{1}{2}$) with (2) the behavior of the system can be investigated, when randomly selected
ferromagnetic interactions are replaced by antiferromagnetic ones of the same strengths or vice versa.

The main feature of the ground state is frustration.
It is shown in \cite{kobe}, \cite{kobeAM}, \cite{koklfr} that for Ising spin glasses a misfit parameter $\mu_0$ can be introduced
as a measure for frustration:
\begin{equation}
\mu_0 = \frac{1}{2} \left ( 1- e_0/e^{id}_{0} \right )  .
\end{equation}
It takes into account that starting from an 'ideal' unfrustrated system with the ground-state energy $E^{id}_{0}$
each unsatisfied bond leads to an increase of the ground-state energy double its
strength \cite{barah82}. Consequently, $e^{id}_{0}$ in (3) is identical to the value for the ferromagnet case ($x = 0$) for all models (2).

The misfit parameter (3) describes the fraction of each bond of the system, which is on average not satisfied in the ground state.
For the antiferromagnetic triangular lattice, for example, the value is $\mu_0 = \frac{1}{3}$, because one
of three bonds of equal strength cannot be satisfied. The same
parameter was used by Stein et al. \cite{stein} to characterize the ground-state
energy of $\pm J$ models in dependence on the concentration of antiferromagnetic
bonds. In a similar way, a related parameter `ground-state energy per bond` is used by Vogel et al. \cite{vogel}.

Obviously, the maximum of $\mu_0$
is $\frac{1}{2}$ for highly frustrated systems (e.g. for high-dimensional
hypercubic and fcc fully frustrated $\pm J$ systems \cite{derrida}, \cite{pincus},
\cite{koklfr}). Because for the SK model $e_0=O(1)$ and $e^{id}_{0}=O(N^{1/2})$ for $N \rightarrow \infty$,
it belongs also to the class of systems with maximum occurring frustration ($\mu_0 = \frac{1}{2}$).
(Erroneously, the correct $N$ dependence of $e^{id}_{0}$ was not taken into account in \cite{koklfr}, so that the discussion about the misfit of the SK model is wrong in that paper.)

\begin{figure}
\includegraphics[angle=-90,scale=0.35]{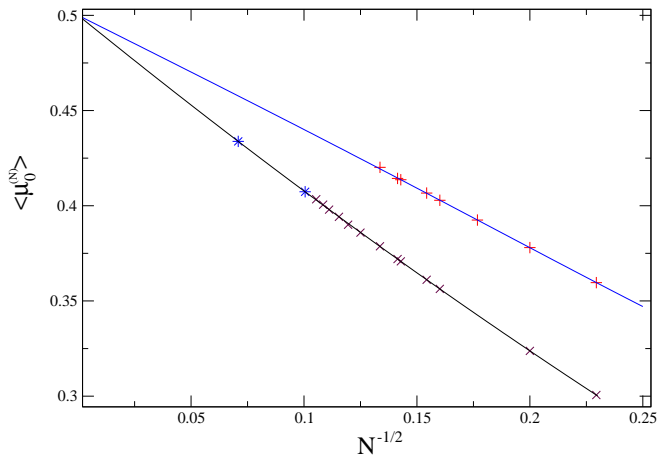}
\caption{\label{fig:epsart} Mean misfit parameter $\langle \mu^{(N)}_0 \rangle$ as a function of $N^{-1/2}$ for the SK model ($\times$) and the fully connected antiferromagnetic system (+); error bars are smaller than symbol sizes. The lines represent the fits on the basis of formula (3), see text. The data points for $N$ = 99 (120,000 samples) and 199 (64,000 samples) ($\star$) are estimated from Palassini's results obtained by a hybrid genetic algorithm \cite{pal}.}
\end{figure}

Exact ground-state energy and misfit parameter were determined by using the branch-and-bound algorithm for the SK model and the fully connected afm model, the sizes studied are $N = 19$ to 90 and 56, respectively. The numbers of samples range from 210,000 ($N=19$) to 34 ($N=90$, SK) and to 99 ($N=56$, afm). The mean misfit parameters $\langle \mu^{(N)}_0 \rangle$ are plotted in Fig. 1 as a function of $N^{-1/2}$ and fitted by (3). For this purpose $e^{id}_0$ in (3) is expanded in ascending powers of $N^{-1/2}$. Assuming that the $N$ dependence of the mean ground-state energy per spin follows $\langle e^{(N)}_0 \rangle = e^{\infty}_0 + bN^{-\omega}$ with $\omega = \frac{2}{3}$, the fit results in $e^{\infty}_0 =  - 0.7637 \pm 0.0004$, which is close to the analytical result ($e^{RSB}_{0} = - 0.76321$ \cite{cris}). A fit with $\omega = 0.671$ shifts $e^{\infty}_{0}$ even closer to $e^{RSB}_{0}$. This tendency was also found in \cite{pal}. A determination of the finite-size scaling of the standard deviation $\sigma = (\langle (e^{(N)}_0)^2 \rangle - \langle e^{(N)}_0 \rangle^2)^{1/2}$ is restricted by the small number of samples for larger $N$. It results in $\rho \simeq 0.72$ assuming $\sigma \propto N^{-\rho}$, which is close to Palassini's result ($\rho \simeq \frac {3}{4}$) \cite{pal}. It is pointed out in \cite{pal} that the true values of $\rho $ is probably slightly larger than $\frac {3}{4}$.

\begin{figure}
\includegraphics[angle=-90,scale=0.35]{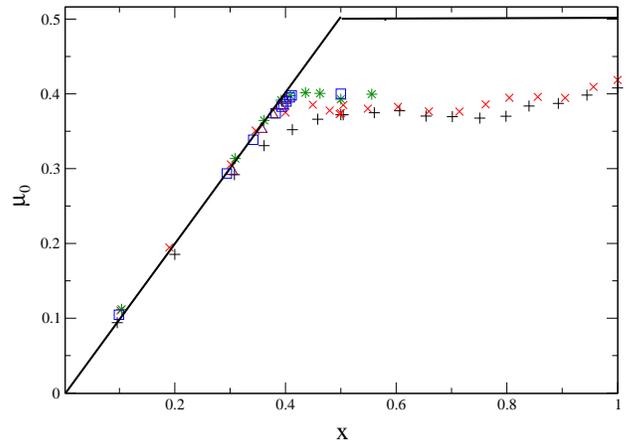}
\caption{\label{fig:epsart} Misfit parameter $\mu_0$ as function of the concentration $x$ of antiferromagnetic bonds for some systems with $N = 49(+) , 56(\times) , 81(\star) , 90(\Box) , 102(\bigtriangleup)$. The $N \rightarrow \infty$ behavior is suggested by the straight line (Equ. (4)).}
\end{figure}

Much less is known about the related afm systems, which are even higher frustrated already for smaller $N$. Preliminary results for $\langle e^{(N)}_0 \rangle$ vs. $N$ seem to exclude the validity of the same $\omega$ value as it is proposed for the SK model. Instead, $\omega = 2$ is chosen leading to $e^{\infty}_{0,\mbox{\tiny {afm}}} =  - 0.474$. The estimation of the $N$ dependence of the standard deviation results in $\rho_{\mbox{\tiny {afm}}} = 0.74 \pm 0.03$.

The misfit parameter depending on the fraction of antiferromagnetic bonds $x$ of the model (2) is shown in Fig. 2 for some finite systems.
For $x \ll 1$ the ground-state energy increases linearly with $x$ according to $e_0(x) = e^{id}_0(1 - 2x)$, i.e. $\mu_0(x) = x$, as long as
the ground state remains ferromagnetically ordered. Obviously, this behavior persists also with increasing $x$, whereas $\mu_0$ is stabilized for $x > 0.5$. (A relatively flat minimum for the finite systems can be understood keeping in mind that the basic element of the system is the triangle configuration, which has the tendency to reduce the frustration, when it is built of two antiferromagnetic and one ferromagnetic bonds. This situation arises relatively often at $x = \frac{2}{3}$.) Otherwise, it can be recognized from the results presented in Fig. 1 that  $\mu_0(x \ge \frac{1}{2}) = \frac{1}{2}$ in the thermodynamic limit $N \rightarrow \infty$. So the numerical results for system sizes up to $N=102$ suggest
\begin{equation}
\mu _0 (x) = \left \{ {x \atop 0.5} \quad \mbox{for} \; \; x \quad { < 0.5 \atop  \geq 0.5} \right.
\end{equation}
and imply that starting from the SK model ($x = \frac{1}{2}$) immediately a ferromagnetic ground state appears, when antiferromagnetic interactions
partly are replaced by ferromagnetic ones.

In summary, exact numerical data for the ground-state energy and its finite-size scaling are presented, which are in agreement with recent results of Palassini \cite{pal} obtained using hybrid genetic algorithm. They also provide the $T \rightarrow 0$ limit of the energy $e_{quench}$ of {\it inherent structures}, i.e. the mean energy of the minima accessible from equilibrium configurations \cite{coluzzi}, \cite{parisifrac}. Moreover, exact results
are important to control the efficiency of approximated algorithms, which can be applied for larger systems \cite{busso}.

The SK model belongs to the class of maximally frustrated systems, which have a misfit parameter $\mu_0 = \frac{1}{2}$.
It is embedded as a special case into related models, which are introduced by varying the signs of the interactions $J_{ij}$. The ground state of
the SK model switches over from a spin-glass state to a ferromagnetic one, when the fraction of the ferromagnetic interactions is larger than the fraction of antiferromagnetic ones. The latter conjecture may also enlighten the controversy on the complexity of the SK model or its vanishing and is challenging for further investigations.

I benefit from discussions and correspondence with A.K. Hartmann, J. Krawczyk, M. Palassini, G. Parisi, T. Plefka, H. Rieger,
M. Tiersch and J. Wei{\ss}barth. The assistance of Ch. Roeper in numerical calculations
is gratefully acknowledged.


\thebibliography{39}

\bibitem[1]{SK}D. Sherrington and S. Kirkpatrick, Phys.\ Rev.\ Lett.\
        {\bf 35}, 1792 (1975); S. Kirkpatrick and D. Sherrington, Phys.\
        Rev.\ B\ {\bf 17}, 4384 (1978).
\bibitem[2]{binyoung}G. Parisi, Phys.\ Rev.\ Lett.\ {\bf 43}, 1754 (1979);
           K. Binder and A. P. Young, Rev.\ Mod.\ Phys. {\bf 58},
           801 (1986);
           M. M\'ezard, G. Parisi, and M. A. Virasoro, {\it
           Spin Glass Theory and Beyond} (World Scientific, Singapore, 1987);
           K. Fischer and J. A. Hertz, {\it Spin Glasses}
           (Cambridge University Press, Cambridge, 1991).
\bibitem[3]{tala}M. Talagrand, Theor.\ Computer\ Science {\bf 265}, 69 (2001).
\bibitem[4]{steinrev}D. L. Stein,  to appear in {\it Quantum Decoherence and Entropy in Complex Systems}, ed. T. Elze (Springer), cond-mat/0301104.
\bibitem[5]{franz}S. Franz and F. L. Toninelli, cond-mat/0310356.
\bibitem[6]{plefka}T. Plefka, cond-mat/0310782.
\bibitem[7]{hed}G. Hed, A. P. Young, and E. Domany, cond-mat/0310609.
\bibitem[8]{hartrie}A. K. Hartmann and H. Rieger, {\it
           Optimization Algorithms in Physics} (Wiley-VCH, Berlin, 2002).
\bibitem[9]{and}A. Andreanov, F. Barbieri and O. C. Martin, cond-mat/0307709.
\bibitem[10]{pal}M. Palassini, cond-mat/0307713.
\bibitem[11]{koha}S. Kobe and A. Hartwig, Computer Phys.\ Commun.\ {\bf 16}, 1 (1978).
\bibitem[12]{hadako}A. Hartwig, F. Daske, and S. Kobe, Computer Phys.\ Commun.\ {\bf 32}, 133 (1984).
\bibitem[13]{korev}S. Kobe, in {\it Computer Simulation Studies in Condensed-Matter Physics IX},
        Springer Proc. in Physics Vol. 82, edited by D. P. Landau, K. K. Mon, and H.-B. Sch\"uttler (Springer-Verlag,
        Berlin, 1997), p. 178; S. Kobe and J. Krawczyk, to appear in {\it Computational Complexity and Statistical Physics},
	edited by A. Percus, G. Istrate, and C. Moore.
\bibitem[14]{kobe}S. Kobe and K. Handrich, phys.\ stat.\ sol.\ (b) {\bf 73},
        K65 (1976)
\bibitem[15]{kobeAM}S. Kobe, in {\it Amorphous Magnetism II}, ed. by R. A.
        Levy and R. Hasegawa (Plenum, New York, 1977), p. 529.
\bibitem[16]{koklfr}S. Kobe and T. Klotz, Phys.\ Rev.\ E\ {\bf 52}, 5660 (1995).
\bibitem[17]{barah82}F. Barahona, J.\ Phys.\ A:\ Math.\ Gen. {\bf 15}, 3241 (1982).
\bibitem[18]{stein}D. L. Stein, G. Baskaran, S. Liang, and M. N. Barber,
        Phys.\ Rev.\ B\ {\bf 36}, 5567 (1987).
\bibitem[19]{vogel} E. E. Vogel, J. Cartes, S. Contreras, W. Lebrecht, and
            J. Villegas, Phys.\ Rev.\ B\ {\bf 49}, 6018 (1994).
\bibitem[20]{derrida}B. Derrida, Y.Pomeau, G. Toulouse, and J. Vannimenus,
        J.\ Phys.\ (Paris)\ {\bf 40}, 617 (1979).
\bibitem[21]{pincus}S. Alexander and P. Pincus, J.\ Phys.\ A:\ Math.\ Gen.\
        {\bf 13}, 263 (1980).
\bibitem[22]{cris}A. Crisanti and T. Rizzo, Phys.\ Rev. E {\bf 65}, 046137 (2002).
\bibitem[23]{coluzzi}B. Coluzzi, E. Marinari, G. Parisi, and H. Rieger, J.\ Phys.\ A:\ Math.\ Gen.\
        {\bf 33}, 3851 (2000).
\bibitem[24]{parisifrac}G. Parisi, Fractals {\bf 11}, Suppl. issue (Febr. 2003), 161 (2003).
\bibitem[25]{busso}L. Bussolari, P. Contucci, M. Degli Esposti, and C. Giardin\a`{a},
            J.\ Phys.\ A:\ Math.\ Gen. {\bf 36}, 2413 (2003).
\end{document}